# Evaluation of Reduced-Graphene-Oxide Aligned with $WO_3$-Nanorods as Support for Pt Nanoparticles during Oxygen Electroreduction in Acid Medium

Iwona A. Rutkowska[a]*, Sylwia Zoladek[a], Anna Wadas[a], Krzysztof Miecznikowski[a], Magdalena Skunik-Nuckowska[a], Enrico Negro[b], Vito Di Noto[b], Agnieszka Zlotorowicz[a,b], Piotr Zelenay[a,c], Pawel J. Kulesza[a]*

[a]*Department of Chemistry, University of Warsaw, Pasteura 1, PL-02-093 Warsaw, Poland*
[b]*Department of Industrial Engineering, Università degli Studi di Padova in Department of Chemical Sciences, Via Marzolo 1, 35131 Padova (PD) Italy*
[c]*Los Alamos National Laboratory, Materials Physics and Applications, Los Alamos, New Mexico 87545, U.S.A.*

*Corresponding Authors: P.J. Kulesza (pkulesza@chem.uw.edu.pl)
I. A. Rutkowska (ilinek@chem.uw.edu.pl)

**Abstract**

Hybrid supports composed of chemically-reduced graphene-oxide-aligned with tungsten oxide nanowires are considered here as active carriers for dispersed platinum with an ultimate goal of producing improved catalysts for electroreduction of oxygen in acid medium. Here $WO_3$ nanostructures are expected to be attached mainly to the edges of graphene thus making the hybrid structure not only highly porous but also capable of preventing graphene stacking and creating numerous sites for the deposition of Pt nanoparticles. Comparison has been made to the analogous systems utilizing neither reduced graphene oxide nor tungsten oxide component. By over-coating the reduced-graphene-oxide support with $WO_3$ nanorods, the electrocatalytic activity of the system toward the reduction of oxygen in acid medium has been enhanced even at the low Pt loading of 30 μg cm$^{-2}$. The RRDE data are consistent with decreased formation of hydrogen peroxide in the presence of $WO_3$. Among important issues are such features of the oxide as porosity, large population of hydroxyl groups, high Broensted acidity, as well as fast electron transfers coupled to unimpeded proton displacements. The conclusions are supported with mechanistic and kinetic studies involving double-potential-step chronocoulometry as an alternative diagnostic tool to rotating ring-disk voltammetry.

**Introduction**

Development of catalytic systems for oxygen reduction reaction (ORR), particularly with respect to potential applications in low-temperature fuel cells [1-13] is still one of the most important areas of electrocatalysis. Considerable research efforts have centered on the development of the Pt-free or low-Pt-content catalytic systems [14]. An ultimate goal would be to replace completely Pt with non-precious metal catalysts, e.g. from the metal-nitrogen/carbon compounds [15-17], metal oxide/oxysalts [18,19], metal-organic-frameworks [20] or just bare or functionalized carbons [21,22]. Although these nonprecious metal alternatives could exhibit in principle activities comparable to, or sometimes even better than, those of Pt in alkaline electrolytes, most of them have been less promising both in terms activity and stability (relative to t-based systems) in acid environments. Indeed, the commercially available applications of fuel cells (e.g. in vehicles) are mostly based on proton exchange membranes because the practical utilization of alkaline membranes is rather limited [23−25]. Consequently, serious research has been focused on improving the catalytic activity of Pt-based catalysts in acid media with less Pt loading. For example, alloying Pt with transition metals has been attempted to modify electronic structure of Pt surfaces thus resulting in improved ORR activity [26,27]. In other words, there is a need of better utilization of catalytic sites and significant lowering of the noble metal loadings.

In addition to large electrochemically active surface areas and the presence of highly dispersed active sites, while exhibiting long-term stability, a useful support should prevent agglomeration of catalytic centers, facilitate oxygen mass transfer and water removal, and assure good electrical conductivity at the electrocatalytic interface. Because of the high specific surface area and excellent thermal, mechanical and electrical properties, graphene-based electrocatalysts have recently been considered. Obviously, special attention has been first devoted to noble-metal-free carbon-based (e.g. heteroatom-doped, iron or cobalt modified, totally-metal-free functionalized or derivatized) nanomaterials including graphene-type systems [28-34]. Different concepts of utilization, including nanostructuring, doping, admixing, preconditioning, modification or functionalization of various graphene-based systems for catalytic electroreduction of oxygen have been explored [35-46].

In the present work, we consider the chemically-reduced-graphene-oxide-supported dispersed Pt nanoparticles that have been intentionally modified with tungsten oxide nanorods as the catalytic system for the electroreduction of oxygen in acid medium (0.5 mol dm$^{-3}$ $H_2SO_4$). Among important issues is the ability of the proposed material is the ability to effectively induce decomposition of the hydrogen peroxide undesirable intermediate. The latter problem is expected to become an issue when the catalytic platinum would be utilized at low loadings. Here, we propose to decorate the graphene-type carriers with Pt nanoparticles (loading, 30 μg cm$^{-2}$) or with tungsten oxide nanorods (loading, 200 μg cm$^{-2}$). The usefulness of $WO_3$ nanostructures during the reduction of oxygen has been recently demonstrated [4]. It is noteworthy that $WO_3$ has been found to exhibit hreactivity toward the reductive decomposition of hydrogen peroxide. Furthermore, we have utilized the so-called reduced graphene oxide which, contrary to conventional graphene, still contains oxygen functional groups regardless of subjecting it to the chemical reduction step [47]. By analogy to graphene oxide, the existence of oxygen groups in the plane of carbon atoms of reduced graphene oxide not only tends to increase the interlayer distance but also makes the layers somewhat hydrophilic. It is apparent from the conventional and rotating ring-disk voltammetric measurements that the systems utilizing Pt nanoparticles, reduced-graphene-oxide-supports decorated with tungsten oxide nanorods behave as the potent $O_2$-reduction electrocatalytic systems. The utility of double-potential-step chronocoulometry as the diagnostic tool has also been demonstrated.

**Experimental**

Chemical reagents were analytical grade materials. Tungsten(VI) oxide nanowires and 5% Nafion-1100 solution were purchased from Aldrich. Platinum black was obtained from Alfa Aesar. Sulfuric acid and formic acid were from POCh (Gliwice, Poland).

All solutions were prepared using doubly-distilled and subsequently deionized (Millipore Milli-Q) water. They were deoxygenated by bubbling with high purity nitrogen. Measurements were made at room temperature (22±2°C).

All electrochemical measurements were performed using a CH Instruments (Austin, TX, USA) Model 760D workstation in three electrodes configuration. The glassy carbon working electrode was utilized in a form of the disk of geometric area, 0.071 cm$^2$. The reference electrode was the $K_2SO_4$-saturated $Hg_2SO_4$ electrode, and the carbon rod was used as the counter electrode. All potentials reported here were recalculated and, unless otherwise stated, expressed *vs.* Reversible Hydrogen Electrode (RHE).

The rotating ring-disk electrode (RRDE) voltammetric experiments were conducted via variable speed rotator (Pine Instruments, USA). RRDE assembly included a glassy carbon disk (diameter 5.61 mm) and a platinum ring (inner and outer diameters were 6.25 and 7.92 mm, respectively) and its collection efficiency was equal to 0.39. Prior the experiments the working electrode was polished with aqueous alumina slurries (grain size 0.05 μm) on a Buehler polishing cloth.

Reduced graphene oxide (rGO) was obtained in the course of the hydrazine reduction method in an analogous manner as described before [37]. In brief, 10 ml of 50 % hydrazine water solution was added to 100 ml of 0.5 wt.% GO water dispersion. The mixture was heated up to 100 ºC and kept under stirring for 2 h. After reduction, the product was filtered using polyethersulfone (PES) filter with 0.8 μm pore size. Reduce graphene oxide (rGO) were prepared by dispersing 50 mg of them using ultrasonic bath in 5 cm$^3$ of deionized water. To modify glassy carbon electrode with carbon materials, the suspension of rGO (2 μdm$^3$) was deposited on the electrode and dried in air. Loading of rGO was equal 300 μg cm$^{-2}$.

Suspensions (inks) of platinum black nanoparticles were prepared by dispersing 2.1 mg of the Pt commercial samples through sonication for 120 min in 2.0 cm$^3$ of distilled water to obtain a homogenous mixture. Later, 2 μdm$^3$ of the ink was placed onto the film surface to yield the noble metal loading of 30 μg cm$^{-2}$.

WO$_3$ nanowires were prepare by dispersing 1.0 mg of was dispersed in 0.1 cm$^3$ of deionized water using an ultrasonic bath for 1 h. Later 2.0 μdm$^3$ of aqueous suspension of WO$_3$ nanowires was used to cover catalytic films. Loading of WO$_3$ nanowires was equal 300 μg cm$^{-2}$.

To prepare hybrid system, the first rGO was dropped on glassy carbon electrode, later Pt nanoparticles and the end ink of tungsten oxide nanowires.

As a rule, the films were over-coated and stabilized with ultrathin layers of Nafion polyelectrolyte by depositing 1 μdm$^3$ of the Nafion solution (prepared by introducing 5%mass of the commercial Nafion solution into ethanol at the 1 to 10 volumetric ratio).

The catalytic materials (films) were first pre-treated in the deaerated 0.5 mol dm$^{-3}$ H$_2$SO$_4$ electrolyte by subjecting them to repetitive potential cycling (in the range from 0.04 to 1.04 V vs. RHE) at 10 mV s$^{-1}$ for 30 min. Before the actual electrocatalytic experiments were performed, the electrodes were conditioned by potential cycling (10 full potential cycles at 10 mV s$^{-1}$) in the potential range from 0.04 to 1.04 V (vs. RHE) in the oxygen-saturated 0.5 mol dm$^{-3}$ H$_2$SO$_4$. Before each representative voltammogram was recorded, the working electrode was kept for 20 s ("quiet time") at the starting potential.

Transmission electron microscopy (TEM) images were obtained with JEM 1400 (JEOL Co., Japan, 2008) equipped with high resolution digital camera (CCD MORADA, SiS-Olympus, Germany).

**Results and Discussion**

*Physicochemical Identity of Graphene-Based and $WO_3$ Nanostructures*

The graphene-based catalytic materials were characterized using Transmission Electron Microscopy (TEM). It is apparent from Fig. 1 that, while the reduced graphene oxide (rGO) deposits are in a form of platelet nanostructures (Fig. 1A), the platinum nanoparticles introduced onto rGO (Fig, 1B) are largely dispersed but in some cases Pt agglomerates are formed. On mechanistic grounds, attachment of Pt may take place at the rGO "defect" sites including surface polar groups. Indeed, it has been established that partially reduced graphene oxide, rGO contain various carbon–oxygen groups (hydroxyl, epoxy, carbonyl, carboxyl), in addition to the large population of water molecules still remaining in the reduced samples. Furthermore, the Raman spectra, which were reported earlier [47], show two large peaks in the range of 1300-1600 $cm^{-1}$: one peak near 1350 $cm^{-1}$, which stands for the D band originating from the amorphous structures of carbon, and the second one close 1580 $cm^{-1}$, which is correlated with the G band, reflects the graphitic structures of carbon. Intensities of G and D bands in rGO, are lower and higher, respectively, and this result implies presence of interfacial defects as well as the fairly low degree of organization of the graphitic structure of rGO.

Figure 2 illustrates cyclic voltammetric behavior of $WO_3$ nanorods (Inset to Fig. 2) investigated as deposit on glassy carbon. Two dominanting sets of peaks appearing at potentials lower than 0.1 V reflect redox transitions of tungsten oxide consistent with the formation of two partially reduced forms of $WO_3$, namely hydrogen tungsten oxide bronzes of the type $H_{x1}WO_3$ and $H_{x2}WO_3$ (that are characterized by fast electron transfers and distribution of charge) [4]. The formation of the latter form is accompanied by generation of lower tungsten oxides, $WO_{3-y}$, as well as by sorption of hydrogen.

*Reduction of $O_2$ at Pt Nanoparticles Deposited onto rGO-based catalysts*

The rGO-supported Pt nanoparticles are obviously less active (larger sizes and lower electrochemically active surface area) than conventional Vulcan-supported Pt during electroreduction of oxygen under conditions of the RRDE voltammetric diagnostic experiments at the comparable loadings (30 µg cm$^{-2}$). On the other hand, the rGO-supported Pt (Fig. 3A) exhibits higher electrocatalytic activity than the same Pt nanoparticles deposited on bare glassy carbon. Apparently rGO facilitates as support better dispersion of Pt catalytic sites. In the present work, we have also considered (Fig. 3B) the rGO-supported Pt nanoparticles further decorated with $WO_3$ nanorods (200 µg cm$^{-2}$). Comparison of the background-subtracted responses clearly shows that $WO_3$-modified rGO-supported Pt nanoparticles show relatively the highest electrocatalytic currents (curve c in Fig. 3C) re (Curve b).

It is also clear from the rotating ring-disk (RRDE) experiments (Fig. 4) that the disk currents have occurred to be higher during the reduction of oxygen at Pt nanoparticles deposited onto the $WO_3$-modified rGO (Fig. 4B) relative to the performance of Pt deposited onto rGO only (Fig. 4A). As expected, the ring currents characteristic of the formation of the undesirable hydrogen peroxide intermediate are lower in a case of the $WO_3$-decorated system (Fig. 4B') relative to the performance of the bare rGO-supported Pt (Fig. 4A'). Fig. 5A illustrates the respective dependencies of limiting currents on the square root of rotation rate. The negative deviation from linearity (which is indicative of kinetic limitations) is much more pronounced in the case of the $WO_3$-free system. Also Koutecky-Levich reciprocal plots (Fig.

5B) are characterized by higher intercepts in the latter case thus indicating the lower heterogeneous rate constant ($2*10^{-1}$ cm s$^{-1}$ vs. $3*10^{-1}$ cm s$^{-1}$ upon addition of $WO_3$ nanorods). Fig. 5C illustrates the percent amount of $H_2O_2$ (%$H_2O_2$) formed during reduction of oxygen under the conditions of RRDE voltammetric experiment and application of 1600 rpm rotation rate.. The actual calculations have been done using the equation given below:

$$\%_{H2O2} = 200 * I_{ring}/N / (I_{disk} + I_{ring} / N) \qquad [1]$$

where $I_{ring}$ and $I_{disk}$ are the ring and disk currents, respectively, and N is the collection efficiency (equal to 0.39). The results clearly show that the production of $H_2O_2$ is lower for system utilizing tungsten oxide nanorods, particularly at potentials lower than 0.4 V.

The overall number of electrons exchanged per $O_2$ molecule (n) was calculated as a function of the potential using the RRDE voltammetric data of Fig. 5C and the equation given below:

$$n = 4 * I_{disk} / (I_{disk} + I_{ring} / N) \qquad [2]$$

The corresponding number of transferred electrons (n) per oxygen molecule (Fig. 5D involved in the oxygen reduction was obviously higher in a case of the system utilizing $WO_3$-decorated Pt nanoparticles supported onto rGO.

*Reduction of $O_2$ under Chronocoulometric Conditions*

An alternative and a very useful mode of recording the electrochemical responses (here the oxygen electroreduction and the possible reoxidation of the hydrogen peroxide intermediate) is to integrate the current and to report charge passed as a function of time [48].

Chronocoulometry offers important advantages including good signal-to-noise ratio because the act of integration smooths random noise on the current transients. Furthermore, it is possible to separate surface phenomena (double layer-charging, surface electrochemistry) from bulk electrochemical responses (e.g. oxygen reduction).

Figure 6A illustrates a representative response for the double-potential-step experiment performed on oxygen reduction using rGO-supported Pt nanoaprticles (the same as for RDE). Similar responses can be recorded for bare Pt nanoparticles as well as for the WO3-decorated system. When the potential is shifted from 1.2 V, where insignificant electrolysis (at least in solution) takes place, to 0.3 V which is sufficiently negative to enforce a diffusion-limited current. By plotting charge (Q) versus square root of time ($t^{1/2}$) the plot of the "diffusional" charge is linear (in Fig. 6B at $t^{1/2}$ larger than 0.15). But the total diffusional charge Q vs. $t^{1/2}$ generally does not pass through the origin, because additional components of Q arise from double-layer charging and from the electroreduction of any surface species (e.g. PtO, oxygen-containing groups on carbon/rGO, etc.) including molecules (e.g. CO) that might be adsorbed at the starting potential. The charges devoted to these processes are passed very quickly when compared to the slow accumulation of the diffusional component; hence they may be included by adding two time-independent terms:

$$Q = 2nF\pi^{1/2}r^2 [D_{app}^{1/2}C_0] / t^{1/2} + Q_{dl} + nFA\Gamma_0 \qquad [3]$$

where $Q_{dl}$ is the capacitive charge and $nFA\Gamma_0$ quantifies the faradaic component given to the reduction of the surface excess, and $\Gamma_0$ (surface concentration, mol/cm$^2$), of adsorbed species. The intercept of Q vs. $t^{1/2}$ is therefore $Q_{dl} + nFA\Gamma_0$. Thus chronocoulometry permits us is to comment on the surface phenomena occurring at the electrocatalytic interface. This will be a subject of our next communication. An approximate value of $nFA\Gamma_0$ can be obtained by comparing the intercept of the Q-$t^{1/2}$ plot obtained for a solution containing oxygen, with the

"instantaneous" charge passed in the same experiment performed with supporting electrolyte only (Inset to Fig. 6B).

The slopes ($Q/t^{1/2}$) of the linear portions of the diffusional (linear) portions of the Q vs. $t^{1/2}$ plots (Fig. 7) can have diagnostic meaning when it comes to the oxygen reduction. The applicable equation is as follows:

$$[Q/t^{1/2}] = 2nF\pi^{1/2}r^2 [D_{O2}^{1/2}C_{O2}] \qquad [4]$$

where r, $D_{O2}$ and $C_{O2}$ stand for the radius, diffusion coefficient, and concentration of redox centers (here oxygen), as well as n is a number electrons (for oxygen ideally 4), and F is Faraday constant (96500 C mol$^{-1}$). For the oxygen saturated 0.5 mol dm$^{-3}$ $H_2SO_4$ at 20$^o$C, where $D_{O2}$ = 1.4*10$^{-5}$ cm$^2$ s$^{-1}$, and $C_{O2}$ = 1.1*10$^{-6}$ mol cm$^{-3}$ (1.1 mM), the value of $Q/t^{1/2}$ = 1.18*10$^{-4}$ C s$^{-1/2}$ (~1.2 *10$^{-4}$ C s$^{-1/2}$). Using the data of Fig. 7, namely the background (electrolyte response) corrected slope, the value of $Q/t^{1/2}$ equal to 1.11 has been obtained. Simple comparison to the theoretic al value of 1.2 *10$^{-4}$ C s$^{-1/2}$ implies that most likely 3.7 electrons, instead of the theoretical 4 electrons, have been involved in the oxygen reduction. Using the same approach to comment on the behavior of the $WO_3$-decorated system (for simplicity not shown here), the value of $Q/t^{1/2}$ equal to 1.14 and n = 3.8 have been determined. On the whole, these values are consistent with the RRDE results (Fig. 5D).

In the above experiments, fairly short pulses (e.g 0.5 s or 1 s) can be used. The pulse width should be simply adjusted to the experimental conditions and the nature of the catalytic system just to obtain well defined linear portion permitting precise determination of the $Q/t^{1/2}$ slope. Application of the relative longer pulse (e.g. 5 s) may allow us estimation of the amount of hydrogen peroxide intermediate. Under the long pulse conditions, virtually complete reduction of oxygen is possible during the forward step in the vicinity of the electrode surface. Immediate switching to the reverse oxidation step permits estimation of the hydrogen peroxide formation. Here maximum charges (Fig. 8) characteristic of both processes

(2-electron oxidation of $H_2O_2$, and approximately 4-electron reduction of $O_2$) should be considered. Remembering that different number of electrons are involved in both processes, we have found that following the potential step from 1.1 tom 0.3 V, approximately 7 and 9% of $H_2O_2$ are produced at the $WO_3$-decorated and $WO_3$-free rGO-supported Pt nanoparticles. Although comparison to the RRDE data (Fig. 5C) is not straightforward because of different experimental conditions, the results of chronocoulometric estimations are consistent with our RRDE voltammetric observations.

**Conclusions**

This study clearly demonstrates that the chemically-reduced graphene-oxide, particuraly when decorated with tungsten oxide nanorods, acts as a robust and activating support for dispersed Pt nanoparticles during electrocatalytic reduction of oxygen in acid medium (0.5 mol dm$^{-3}$ $H_2SO_4$). For the same loading of catalytic Pt nanoparticles (30 μg cm$^{-2}$), decoration with $WO_3$ results in the formation of lower amounts of the undesirable $H_2O_2$ intermediate. Moreover the onset potential for the oxygen reduction has been the most positive in a case of the system utilizing $WO_3$-decorated reduced graphene oxide. Synergistic effects and activating interactions between catalytic metal nanoparticles, tungsten oxide and nanostructured graphene supports cannot be excluded here with respect to lowering the dissociation activation energy for molecular $O_2$ through accelerating the charge transfer from metal in the presence of graphene and by reducing stability of the $H_2O_2$ intermediate species.

We have also demonstrated the usefulness of the double potential step chronocoulometry, particularly the charge vs. square root of time (so called Anson) plots as the fast and reliable diagnostic tool for evaluation of the effectiveness of the oxygen reduction.

**Acknowledgements**

We acknowledge the European Commission through the Graphene Flagship – Core 1 project [Grant number GA-696656] and Maestro Project [2012/04/A/ST4/00287 (National Science Center, Poland)].

**Figure Captions**

**Fig. 1.** Transmission electron micrographs of (A) reduced graphene oxide (rGO) and (B) reduced graphene oxide supported platinum nanoparticles

**Fig. 2.** Cyclic voltammetry of $WO_3$ nanowires (deposited on the glassy carbon disk) in acid medium. Inset: Transmission electron micrograph of $WO_3$ nanowires.

**Fig. 3.** Voltammetric reduction of oxygen at (A) rGO-supported Pt nanoparticles and (B) rGO- supported Pt nanoparticles modified with $WO_3$ nanowires. (C) Background-subtracted linear scan voltammetric responses recorded for the reduction of oxygen at (a) platinum nanoparticles, (b) (rGO-supported platinum nanoparticles and (c) rGO-supported platinum nanoparticles modified with $WO_3$ nanowires Electrolyte: oxygen-saturated 0.5 mol dm$^{-3}$ $H_2SO_4$. Scan rate: 10 mV s$^{-1}$.

**Fig. 4.** Normalized rotating ring-disk voltammograms for oxygen reduction at (A) rGO-supported Pt nanopartices, (B) rGO-supported Pt nanopartices modified with $WO_3$. Electrolyte: oxygen-saturated 0.5 mol dm$^{-3}$ $H_2SO_4$. Scan rate: 10 mV s$^{-1}$. Ring currents are recorded upon application of 1.28 V.

**Fig. 5.** (A) Levich plots and (B) Koutecky-Levich reciprocal plots of current densities versus square root of rotation rate ($\omega^{1/2}$) for the electroreduction of oxygen (at 0.2V) at (black points) rGO-supported Pt nanoparticles and (red points) rGO-supported Pt nanoparticles modified with $WO_3$. (C) Percent fraction of hydrogen peroxide and (D) number of exchanged electrons during reduction of $O_2$ at (black line) at rGO-supported Pt nanopartices, (red line) at rGO - supported Pt nanopartices modified with $WO_3$.

**Fig. 6**. (A) Double potential-step hronocoulometric and (B) chronocoulometric Anson plots response for rGO-supported platinum nanoparticles in oxygen-saturated 0.5 mol dm$^{-3}$ H$_2$SO$_4$. Potential steps from 1.2 to 0.3 V vs. RHE. Pulse width: 0.5s.

**Fig. 7**. Chronocoulometric responses for rGO-supported platinum nanoparticles in (A) 0.5 mol dm$^{-3}$ H$_2$SO$_4$ and (B) oxygen-saturated 0.5 mol dm$^{-3}$ H$_2$SO$_4$. Potential steps from 1.0 to 0.5 V vs. RHE. Pulse width: 0.5s.

**Fig. 8.** Background-subtracted long-pulse double-potential-step chronocoulometric (pulse width, 5s) responses recorded for rGO-supported platinum nanoparticles in oxygen-saturated 0.5 mol dm$^{-3}$ H$_2$SO$_4$. Potential steps from 1.1 to 0.3 V vs. RHE.

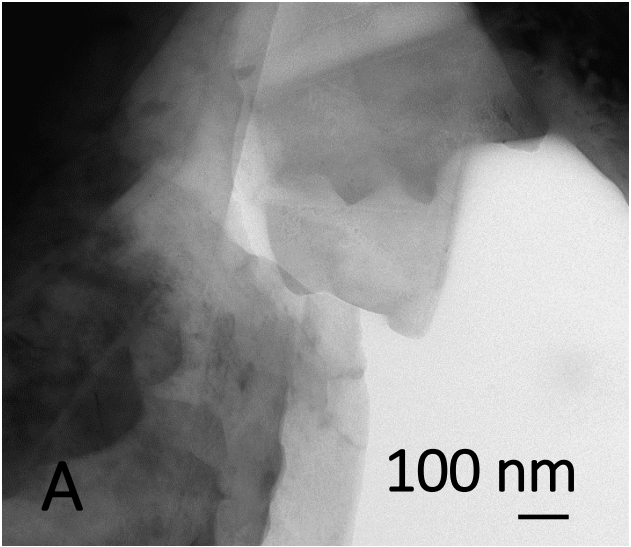
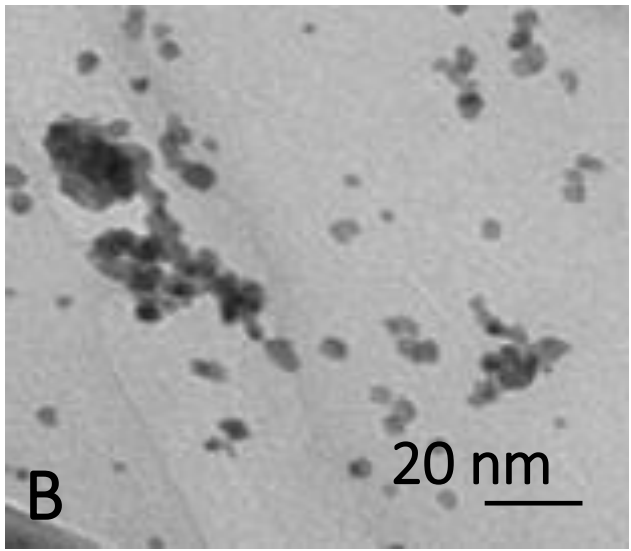

Fig.1

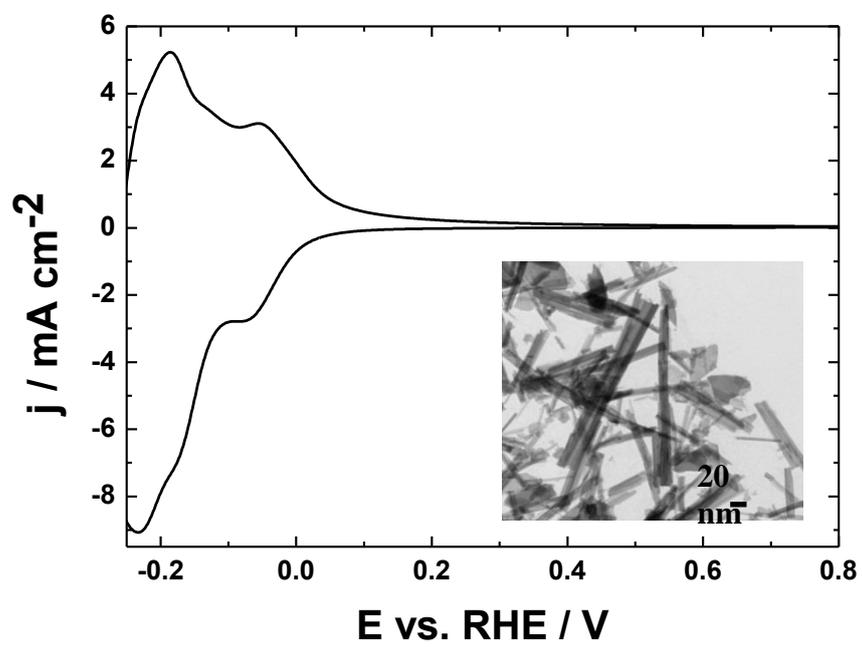

Fig. 2

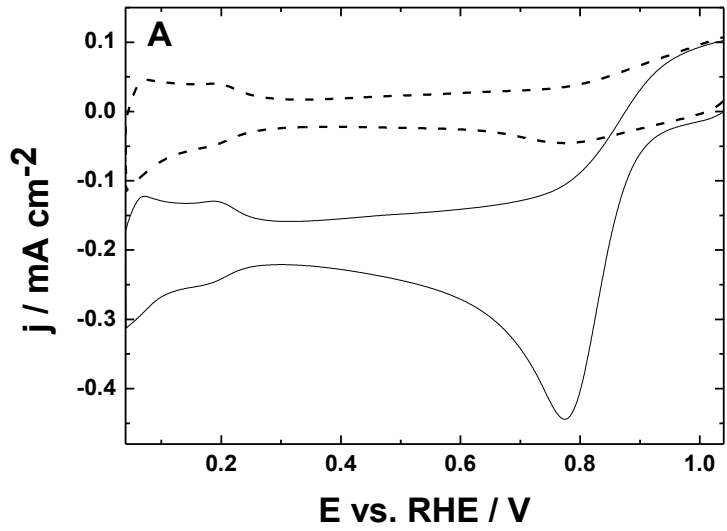

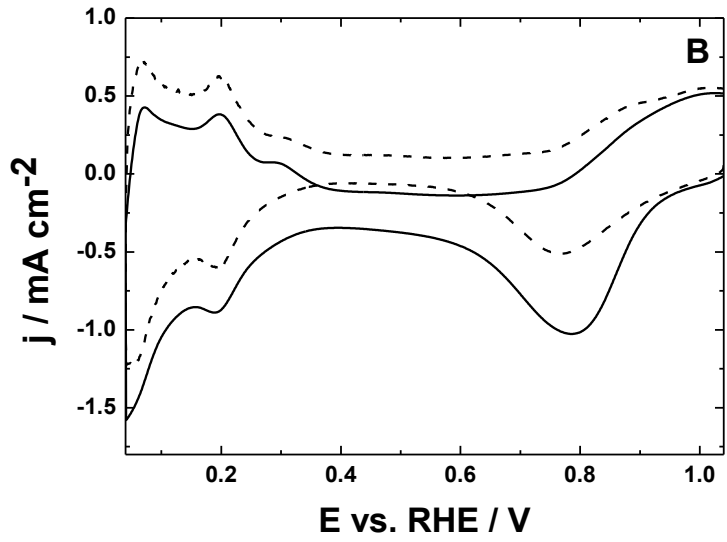

Fig.3

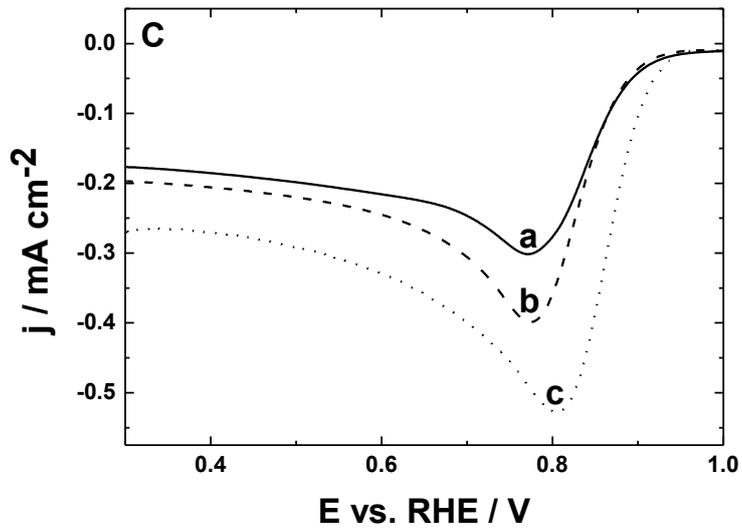

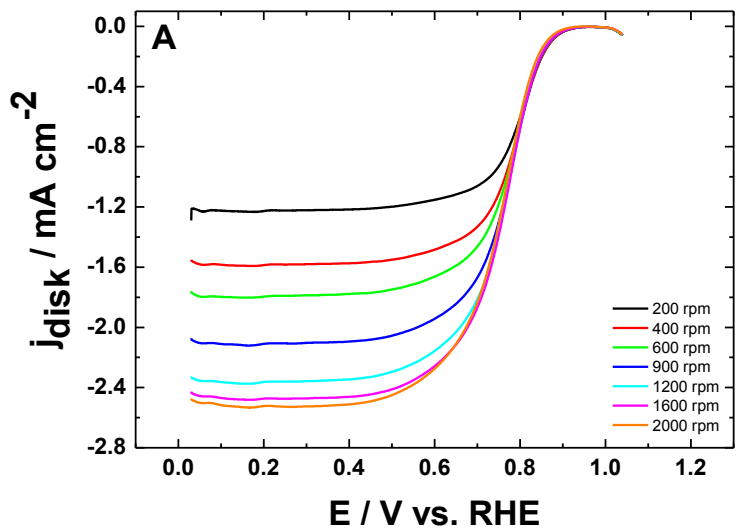
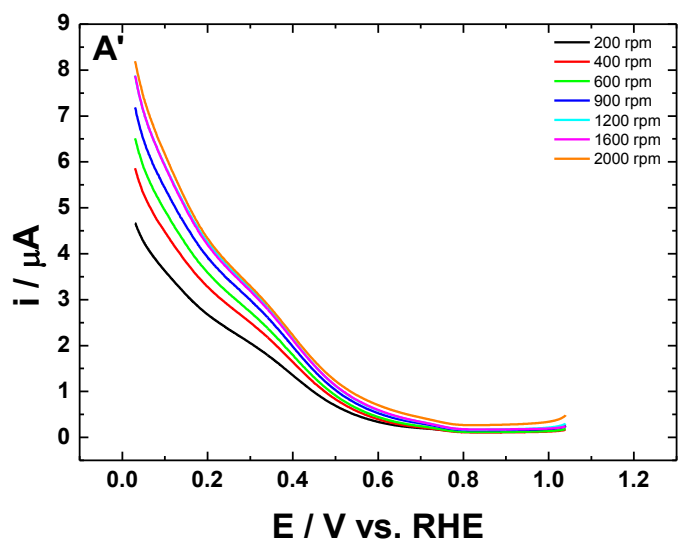

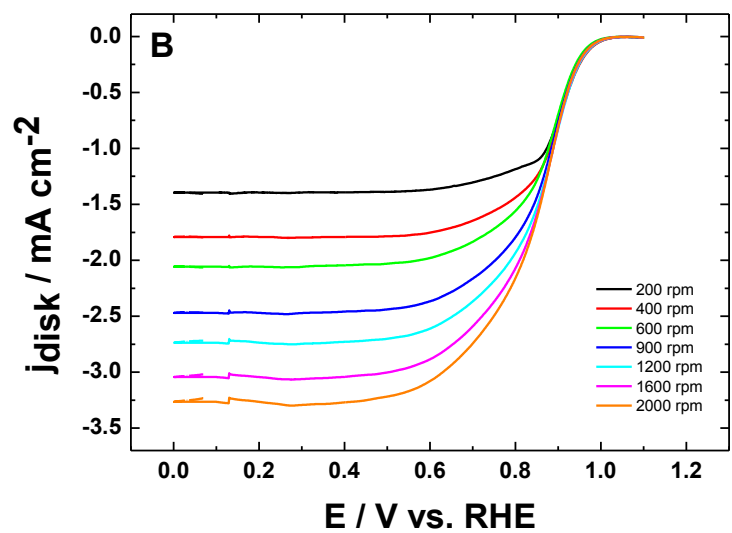

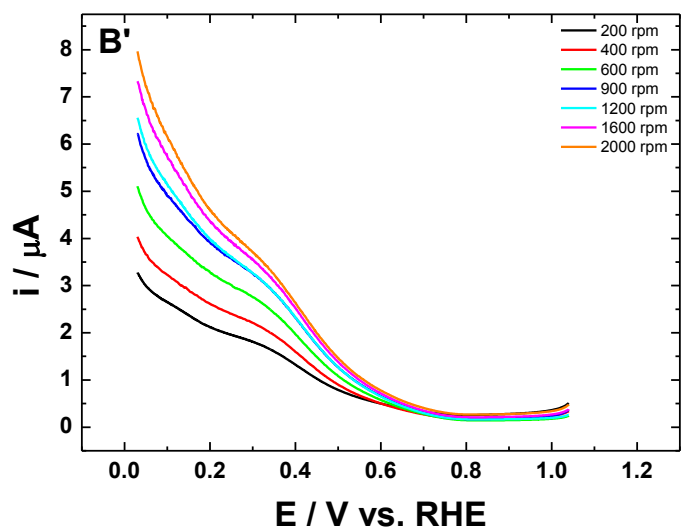

**Fig. 4**

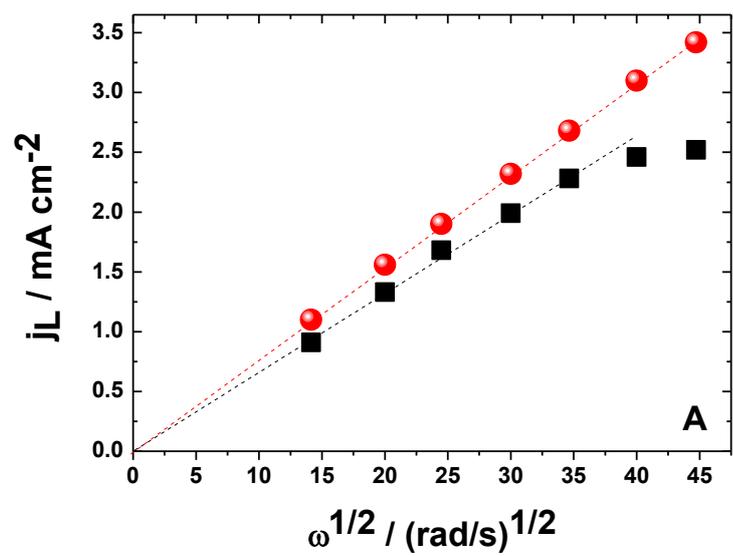

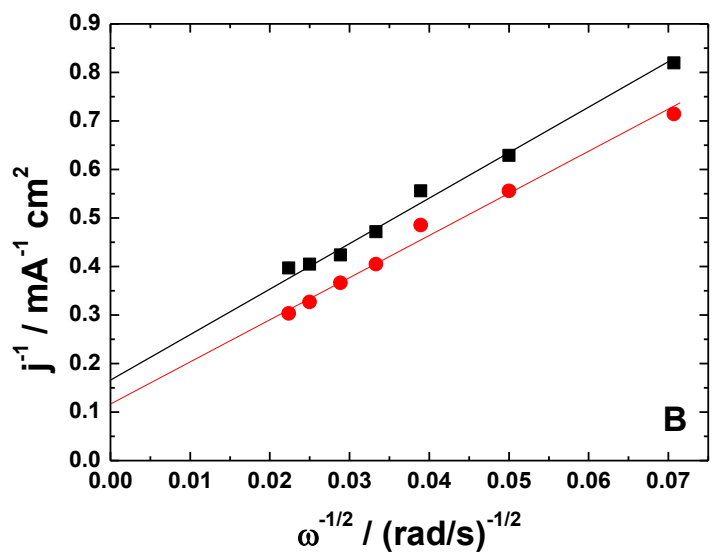

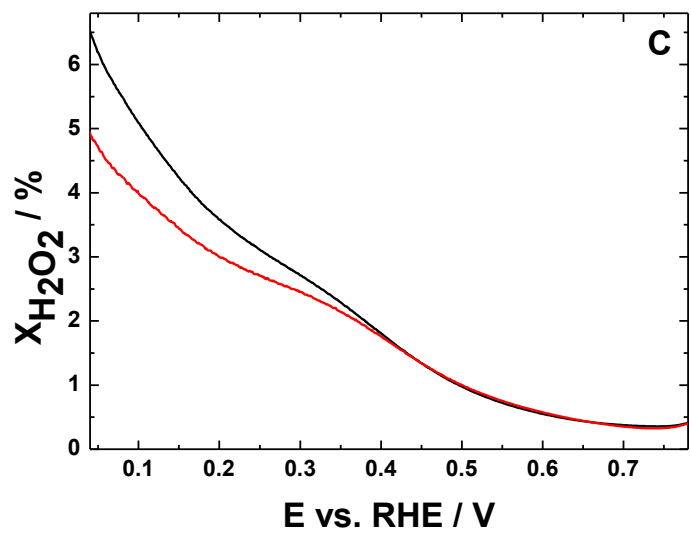

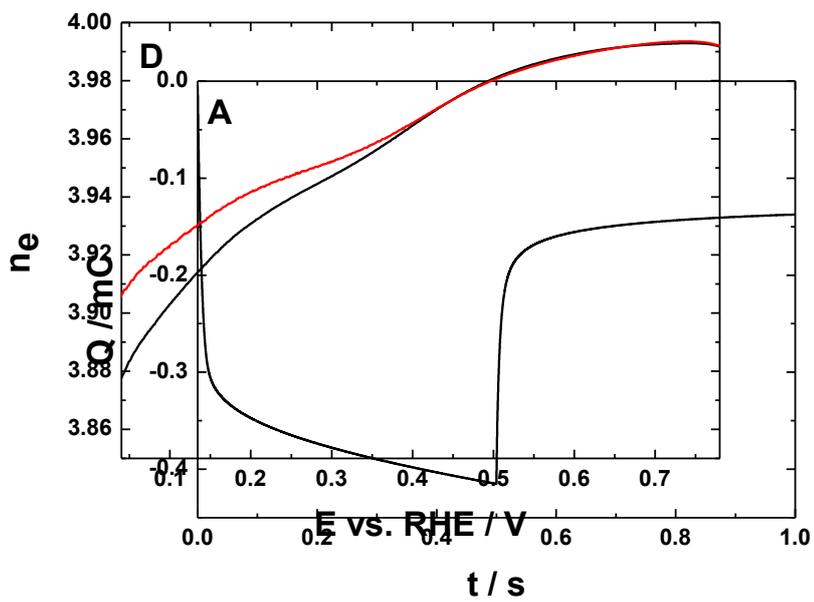

Fig. 5

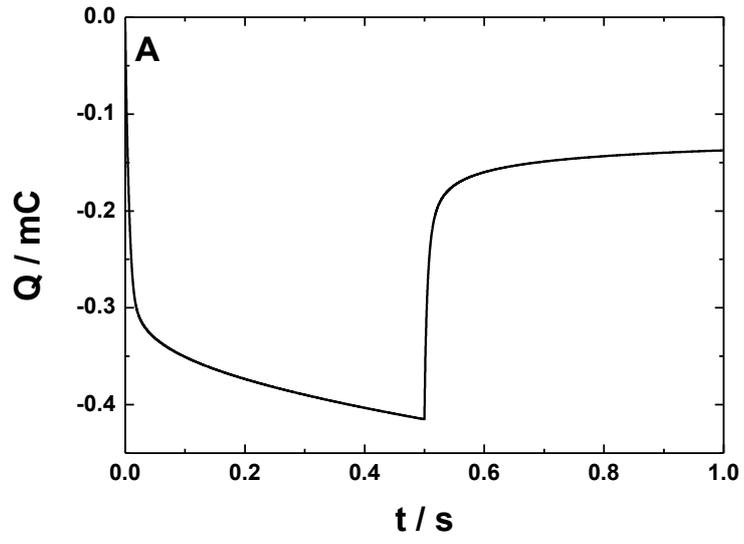

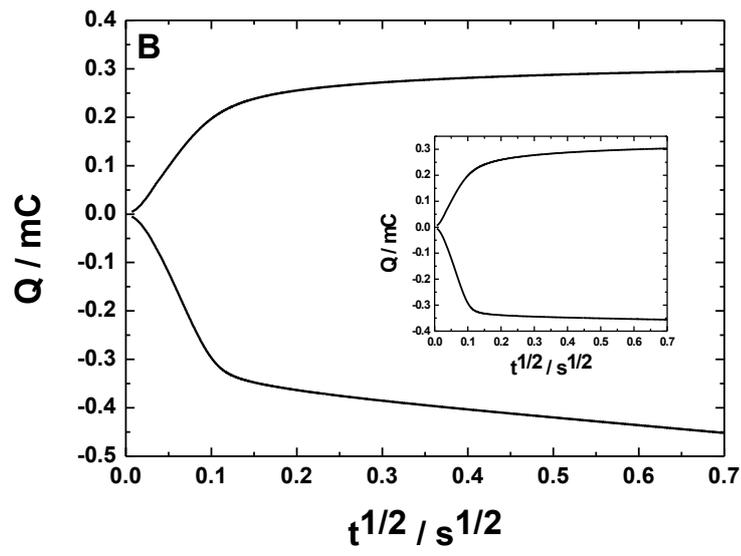

**Fig. 6**

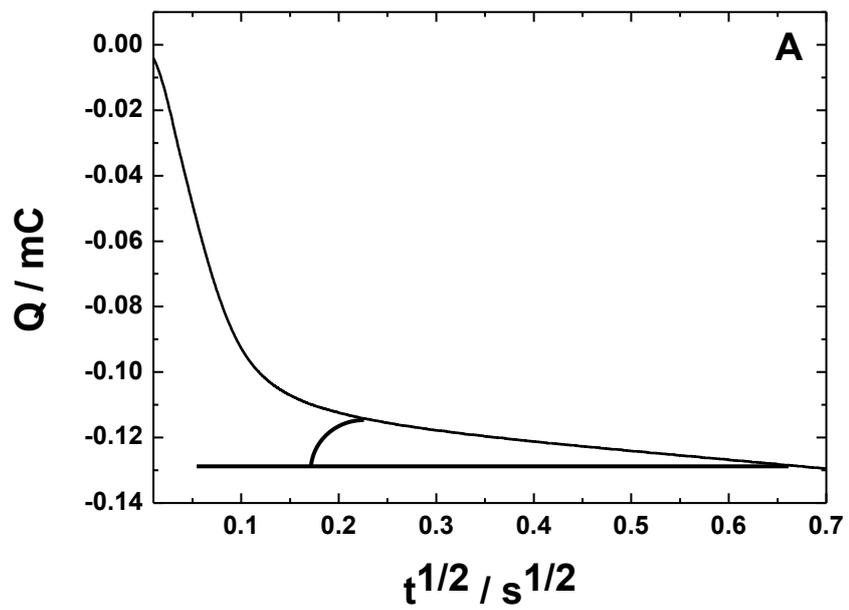

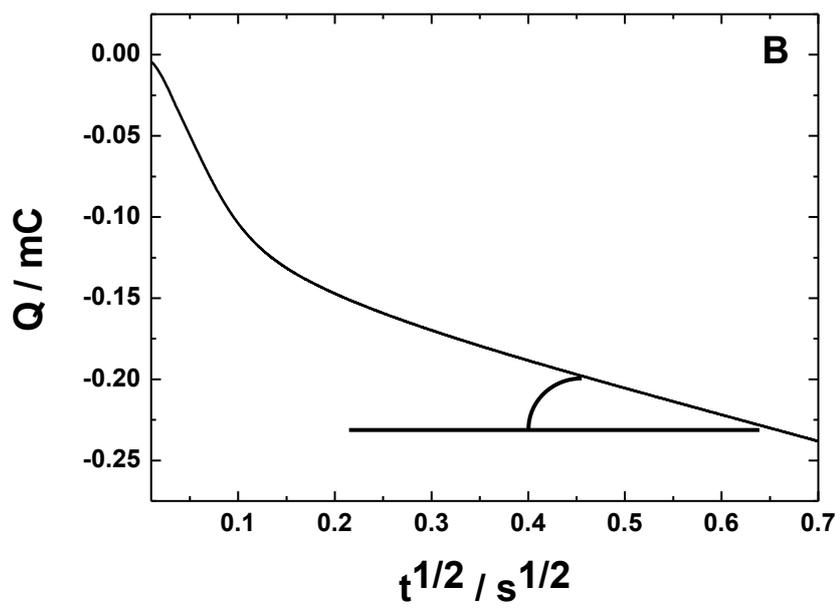

**Fig. 7**

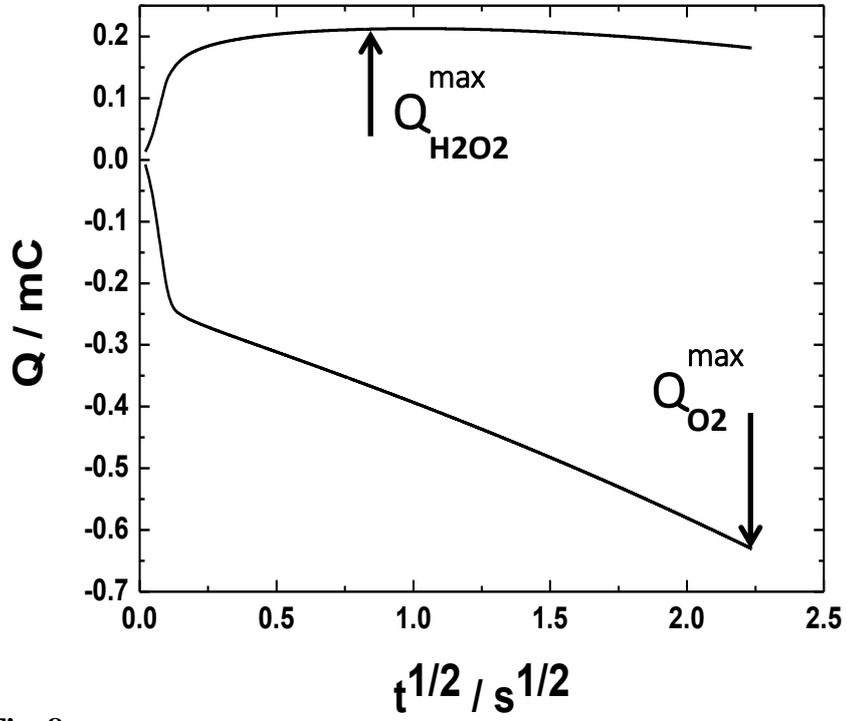

Fig. 8